%% file: 4.15.13.tex
\documentclass[showpacs,preprint,amssymb]{revtex4}
\usepackage{graphicx}
\usepackage{dcolumn}
\usepackage{bm}

\def\HQV{half quantum vortex }

\input mma_defs.tex
\usepackage{amsfonts}
\usepackage{amssymb}

\begin{document}

\title{Numerical study of the stability regions for  half-quantum vortices in superconducting Sr$_2$RuO$_4$}

\author{ KEVIN ROBERTS}

\affiliation{University of Illinois, Department of Physics\\ 1110 W. Green St.\\
Urbana, IL 61801 USA\\E-mail: krobert7@illinois.edu}   

\author{ RAFFI BUDAKIAN}

\affiliation{University of Illinois, Department of Physics\\ 1110 W. Green St.\\
Urbana, IL 61801 USA\\E-mail: budakian@illinois.edu}   

\author{ MICHAEL STONE}

\affiliation{University of Illinois, Department of Physics\\ 1110 W. Green St.\\
Urbana, IL 61801 USA\\E-mail: m-stone5@illinois.edu}   

\begin{abstract}  We  numerically solve the coupled   Landau-Ginzburg-Maxwell equations for a model of  a spin triplet $p_x+ip_y$ superconductor in which   whole or half-quanta   of flux thread through a hole.   We   recover the pattern of stable and unstable  regions for the half-flux quanta observed in a recent experiment.

\end{abstract}

\pacs{74.20.De,74.20.Rp,74.25.Ha,74.70.Pq}

\maketitle

\section{Introduction}

 Superconductors with triplet $p_x+ip_y$ pairing are  interesting  because they can host half-quantum vortices with Majorana core states and non-Abelian braid statistics \cite{ivanov,stern,stone-chung}.   There is indirect evidence that the layered superconductor Sr$_2$RuO$_4$ has this pairing \cite{maeno_review}, and a search is on  for ``smoking gun'' signatures that will confirm this.  One  signature --- chiral edge currents --- has proved elusive, but recent work  \cite{budakian} has found striking results   suggesting that half-quantum vortices have been detected.  If this result is correct, it strongly supports the spin triplet pairing character of the superconducting order parameter.

\begin{figure}[htb]
\centering
\includegraphics{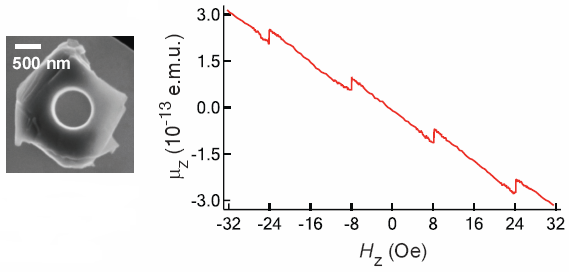}
\caption{SRO sample and magnetization curves showing integer flux transitions adapted from \cite{budakian}.}
\label{fig:ringAndMoment}
\end{figure}

 In the   experiment  reported in \cite{budakian} a micron-sized annular flake  of Sr$_2$RuO$_4$ is 
mounted on a cantilever. Its magnetic moment is monitored as magnetic fields both perpendicular ($B_x$) and parallel ($B_z$) to the c-axis are applied.    As $B_z$ is increased,  moment jumps corresponding to the entry of single-flux-quantum   vortices into the hole  in  the annulus are  easily observed  (see figure \ref{fig:ringAndMoment}).

\begin{figure}[h]
\centering
\includegraphics[width=0.6\linewidth]{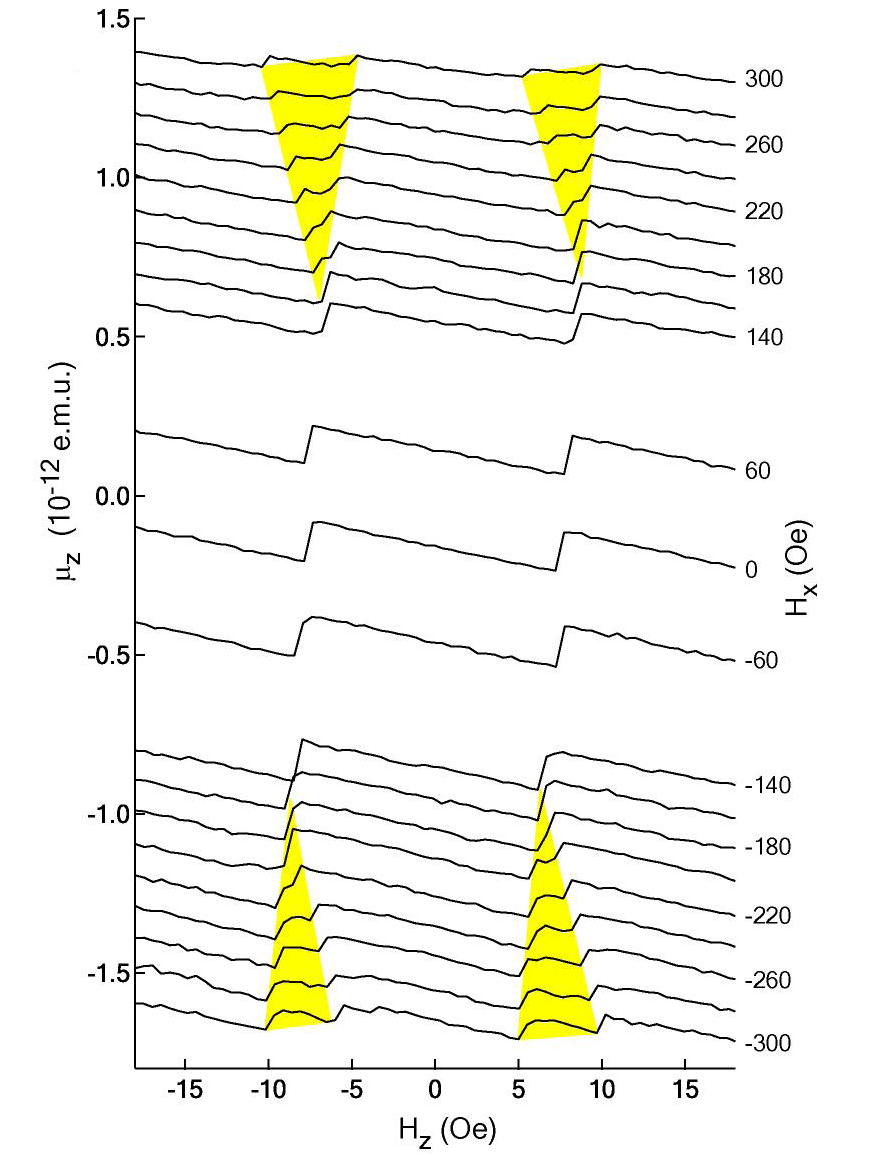}
\caption{Magnetization curves for various values of applied in-plane field.  Adapted from \cite{budakian}. The shaded region highlights the wedge shaped character of the half-flux state's stability region.}
\label{fig:momentData}
\end{figure}

\noindent When a sufficiently large in-plane field $B_x$ is applied, these entry-event jumps break up into two separate events, each  with  one-half of the original magnetic-moment jump.  (See Figure \ref{fig:momentData}.)

The most obvious  explanation is that we are seeing half-quantum vortices: The  large   $B$ field  presumably has broken  a spin 
orbit coupling that had held the spin-triplet  order-parameter ${\bf d}$ vector  parallel to the pair angular momentum vector ${\bf l}$. The  ${\bf d}$ vector    can now  rotate freely in the $x$-$z$ plane and this  freedom permits  the existence of a half-quantum vortex defect  around which the   ${\bf d}$ vector and the order-parameter phase $\phi$  both rotate   through an angle  $\pi$ while  leaving the spin-triplet order parameter single valued. These rotations have the  effect of producing a  phase-winding vortex  in one spin component, while leaving the other with no phase  winding.   A circulating ``anti-vortex'' current is nonetheless induced in the second spin component by the magnetic field, and at large distance the spin-up     and spin-down   flow velocities  become equal and opposite.  This means that  there is  a long-range spin current surrounding the vortex. It is the necessity of reducing the logarithmically divergent free-energy-cost of   this spin current that mandates  the use of small  annular samples \cite{chung-bluhm-kim}.   

As $B_x$ is increased, the separation between the half-quantum jumps becomes larger.  Vakaryuk and Leggett  \cite{vakaryuk-leggett} have proposed  that  this phenomenon can be explained by a {\it kinematic spin polarization\/}.  The different flow velocities for the up and down spin condensates  give rise to  an analogue  of the  Bernoulli effect  which  increases the magnitude of the order parameter for one condensate and decreases it for the other. The resulting magnetic polarization lies parallel to  $B_x$,  and so alters the energy cost for a vortex to enter the sample.   Assessing  whether or not the energy gain from   Vakaryuk-Leggett  mechanism   is sufficient to explain the experimental data requires  a detailed accounting of the energy in the three dimensional magnetic field surrounding the sample  and in the degree of polarization of the condensate. In this paper we perform  this accounting by  obtaining   numerical  solutions to  an appropriate  set of coupled Maxwell-Landau-Ginzburg-equations.  With the  specific  geometry of the samples used in the experiment, and with  reasonable values of the Landau-Ginzburg parameters, we find that   we are   able to qualitively reproduce the experimental data.

In section \ref{SEC:Landau-Ginzburg} we assemble   the ingredients of the two-component Landau-Ginzburg free-energy functional  that we use to model  the  superconductivity. In  section \ref{SEC:numerical-method} we explain how we find the field configurations that minimize our free energy functional, and how  we extract from them   magnetization curves that we can compare with the experimental data.  Section \ref{SEC:results} presents the numerical results, and, finally, section \ref{SEC:discussion} contains a brief discussion of these results and their significance.

\section{Two-component Landau-Ginzburg formalism}
\label{SEC:Landau-Ginzburg}
\subsection{Spin and charge stiffness}

For a  spin-triplet $p_x+ip_y$ superconductor with a fixed  orbital angular momentum vector ${\bf l}=\hat {\bf z}$, the order parameter  is  matrix-valued and of the form   
\be
\left[\matrix{ \Delta_{\uparrow \uparrow} &  \Delta_{\uparrow \downarrow}\cr \Delta_{\downarrow \uparrow} & \Delta_{\downarrow \downarrow}}\right] = |\Delta| e^{i\chi} (-i\sigma_2 \,{\bm \sigma}\cdot {\bf d}) = |\Delta| e^{i\chi} \left[\matrix{ -(d_1+id_2)  &  d_3\cr d_3& d_1-id_2}\right] .
\ee
The ${\bf d}$ vector has unit length, and for our application we will assume that it is perpendicular to a  spin-quantization axis   ${\bf e_3}$,  which need not be the $z$ axis.   We therefore set $d_3=0$, $d_1+id_2=e^{i\phi}$, and  define the phases of $\Delta_{\uparrow \uparrow}$ and $\Delta_{\downarrow \downarrow}$ to be  $\theta_\uparrow+\pi$ and  $\theta_\downarrow$ respectively. Then 
\bea
 \chi&=& \frac12(\theta_\uparrow+\theta_\downarrow),\\
 \phi&=&  \frac12 (\theta_\uparrow-\theta_\downarrow).
\eea 
The authors of \cite{chung-bluhm-kim}  write  the free-energy density in the London form
\be
K_{\rm london}=   \rho_s\left | \nabla \chi-\frac{2e}{\hbar} {\bf A}\right |^2 +\rho_{\rm spin}| \nabla \phi|^2. 
\label{EQ:london}
\ee
This expression    contains only the Goldstone fields $\chi$ and $\phi$ and  so ignores the free-energy cost of gradients in  the magnitude of the order parameter.  Nonetheless  (\ref{EQ:london})  captures the essential far-from-core vortex energetics.  In particular, in a half-quantum vortex either $\theta_\uparrow$ or $\theta_\downarrow$ (but not both) rotate through $\pm 2\pi$. Then $\chi$ and $\phi$  rotate through $\pm \pi$.  Far from the vortex core  the ${\bf B}$  field will adjust so as to make the first term in $K$  zero, so   the total flux threading  the half-vortex is given by
\be
\Phi_{1/2}= \oint {\bf A}\cdot d{\bf r} = \frac{\hbar}{2e} \oint \nabla \chi\cdot d{\bf r} = \frac 14 \left(\frac{2\pi \hbar}{e}\right)=\frac 12 \Phi_0.
 \ee
 The remaining term  {\it cannot\/}  be screened by the ${\bf B}$ field and gives a contribution to the vortex energy that is logarithmically divergent at large distance. 
This divergent energy  cost means that  a half-quantum vortex  be  disfavoured unless  the spin stiffness $\rho_{\rm spin}$ is  small  and   a  finite size to  the superconducting  region cuts-off the long-distance  contribution.

The angle-valued Goldstone fields are not suitable for numerical work as they are not singled-valued in the presence of vortices. 
We need to write the   free energy in terms   single-valued fields.  We therefore introduce   fields 
$\psi_\uparrow =  |\psi_\uparrow |\exp\{i\theta_\uparrow\}$ and  $\psi_\downarrow = |\psi_\downarrow | \exp\{i\theta_{\downarrow}\}$.
The simplest form for the Landau-Ginzburg free-energy density  that has the correct symmetries   contains the kinetic-energy density
\be
\label{eq:CCcoupling}
K_{\rm landau} = \frac{\hbar^2}{2m^*} \left\{\left| \left( \nabla -\frac{2ie}{\hbar}{\bf A}\right)\psi_\uparrow\right|^2+  \left| \left (\nabla -\frac{2ie}{\hbar}{\bf A}\right)\psi_\downarrow\right|^2+ 2b {\bf J}_{\uparrow}\cdot  {\bf J}_{\downarrow}\right\},
\ee
where
\bea 
 {\bf J}_{\uparrow}&=&\frac{ ie\hbar}{m^*}\left(\psi_\uparrow^*\left(\nabla-\frac{2ie}{\hbar}{\bf A}\right) \psi _\uparrow- \psi_\uparrow\left( \nabla+\frac{2ie}{\hbar} {\bf A}\right)\psi^*_\uparrow\right),\\
 {\bf J}_{\downarrow}&=&\frac{ ie\hbar}{m^*}  \left(\psi_\downarrow^*\left(\nabla-\frac{2ie}{\hbar}{\bf A}\right)) \psi _\downarrow- \psi_\downarrow \left(\nabla+\frac{2ie}{\hbar}{\bf A}\right)\psi^*_\downarrow\right).
\eea
The current-current interaction introduces no new magnitude-gradient free-energy cost, and so does not affect the coherence length. 

If we set $\psi_\uparrow= |\psi| e^{i\theta_\uparrow}$, {\it etc.\/},  and temporarily ignore derivatives of the common magnitude $|\psi|$ , then 
\bea
 &K_{\rm landau} \approx&|\psi|^2 \frac{\hbar^2}{2m^*} \left\{\left|\nabla \theta_\uparrow-\frac{2e}{\hbar}{\bf A}\right|^2 + \left|\nabla \theta_\downarrow-\frac{2e}{\hbar} {\bf A}\right|^2 +2b \left(\nabla \theta_\uparrow-\frac{2e}{\hbar}{\bf A}\right)\cdot \left(\nabla \theta_\downarrow-\frac{2e}{\hbar}{\bf A}\right)\right\}\nonumber\\
 &=& |\psi|^2  \frac{\hbar^2}{2m^*}\left\{2 (1+b)\left| \nabla \chi-\frac{2e}{\hbar}{\bf A}\right|^2 +2 (1-b) | \nabla \phi|^2\right\}, 
 \eea
which is to be compared with  desired London form of \cite{chung-bluhm-kim}.  As $\rho_{\rm spin}$  must be positive,  $b$ must be less than unity. Being just less than  unity encourages half-quantum vortices.

The presence  the term $2b {\bf J}_{\uparrow}\cdot  {\bf J}_{\downarrow}$  in the free energy density alters the  
current ${\bf J}_{\rm maxwell }$ that couples to the   magnetic field. We have  
\be
{\bf J}_{\rm maxwell }=  {\bf J}_\uparrow+  {\bf J}_\downarrow -\frac{8e^2b}{m^*}(|\psi_\uparrow|^2 {\bf J}_\downarrow+ |\psi_\downarrow|^2 {\bf J}_\uparrow).
\ee

\subsection{Kinematic spin polarization}

 Legget and Vakaryuk propose \cite{vakaryuk-leggett}  that the dependence of the \HQV stability region on the applied in-plane field $B_x$ can be understood via the existence of a spontaneous spin polarization in the \HQV state that arises from the difference between the spin-up and spin-down condensate  velocities  ${\bf v}_\uparrow$ and ${\bf v}_\downarrow$  in the half-quantum vortex.    
 
 The polarization occurs    because for each of the two spin components in the Landau-Ginzburg theory we have a version of  Bernoulli's equation:
 \be
 \frac 12 m^*|{\bf v}_{\uparrow,\downarrow}|^2 + \frac{\partial u}{\partial \rho_{\uparrow,\downarrow}}=\hbox{const.}
 \ee 
 Here $\rho_{\uparrow,\downarrow}=|\psi_{\uparrow,\downarrow}|^2$, and 
 \be
 u(\rho)= \alpha \rho +\frac 12 \beta \rho^2,
 \ee 
 is the potential part of the Landau-Ginzburg free energy density.
 
 As a consequence,  the  faster the superflow, the lower the  order-parameter density.
 The polarizing tendency  is proportional to
 \bea
 |{\bf v}_\uparrow|^2-  |{\bf v}_\downarrow|^2&=& ({\bf v}_\uparrow+{\bf v}_\downarrow)\cdot({\bf v}_\uparrow-{\bf v}_\downarrow)\\
 &=& {\bf v}_{\rm charge}\cdot {\bf v}_{\rm spin}
 \eea
 This  polarizing tendency gives rise to  spin magnetic moment 
\be
{\bm  \mu}_{\rm spin}=  g \mu_B \left(|\psi_\uparrow|^2-|\psi_\downarrow|^2\right) {\bf e}_3
  \ee
  Here $\mu_B$ denotes  the  Bohr magneton, and $g$ is a phenomenological parameter that we expect to be of order unity. (If $|\psi|^2_{\uparrow,\downarrow}$     were the actual density of cooper pairs, and  if each of the two electrons in the pair contributes  a Dirac moment of $\mu_{\rm electron}\approx   \mu_B$, we would have $g=2$.)
  The induced moment couples to  the magnetic field to give a free-energy contribution
  \be
  \Delta F=  -{\bf B}\cdot {\bm \mu}_{\rm spin}.
  \ee
 Jang {\it et al.\/}\ \cite{budakian}  assume that this moment lies in the $x$-$y$ plane and so it is affected only by $B_x$.  We  therefore account for it by including a term
  \be \label{eq:KSP}
    \Delta F = - g\mu_B \left(|\psi_\uparrow|^2-|\psi_\downarrow|^2\right) |B_{\parallel}|
    \ee
    in our free-energy functional.    
 
 \subsection{Anisotropy}
 
The superconductor  Sr$_2$RuO$_4$ is quite anisotropic, and it is necessary to  replace the usual Landau-Ginzburg scalar mass $m^*$
with a mass tensor $M^*$ so that the  free energy
becomes
\bea
 F[\psi,\psi^*,{\bf A}]&= &\int _\Omega d^3x\left\{ \frac {\hbar^2}{2}\left(\nabla +\frac{2ie}{\hbar}{\bf A}\right)\psi^* \cdot (M^{*})^{-1}\cdot\left(\nabla -\frac{2ie}{\hbar}{\bf A}\right)\psi+ \right. \nonumber \\
  &&\qquad\qquad\qquad\left.+\alpha |\psi|^2 +\frac{\beta}{2} |\psi|^4 +\frac 1{2\mu_0}|\nabla\times {\bf A}-{\bf B}_{\rm ext}|^2\right\}.
\eea
We choose a coordinate system such that the crystal $ab$ planes lie in the $xy$-plane and are displaced from each other in the $z$-direction. Then, the effective mass tensor takes the form
\be
 M^*= \text{diag}(m_{\parallel},m_{\parallel},m_{\perp})
\ee
where the ratio
\be
\frac{m_{\perp}}{m_{\parallel}}=\left(\frac{\lambda_{\perp}}{\lambda_{\parallel}}\right)^2
\ee 
can be calculated from  the  measured penetration depths $\lambda_\perp$ and $\lambda_\parallel$
of Sr$_2$RuO$_4$ and is found to be approximately 400.

\section{Numerical Method}
\label{SEC:numerical-method}

Our goal is  to create equilibrium magnetization curves that can be directly compared with those in   \cite{budakian}  (see figures \ref{fig:ringAndMoment} and \ref{fig:momentData}). Because of the asymmetric and three dimensional character of the experimental samples, these curves have to be found numerically. 
 
For the numerical calculation, it is convenient to express the  Landau-Ginzburg  free energy  in dimensionless form as 
\bea
\label{EQ:finalFreeEnergy}
F[\psi_{\uparrow},\psi_{\downarrow},{\bf A}]= \int _\Omega d^3x\left\{ \sum_{\uparrow,\downarrow}\left[\left(\frac \nabla {\kappa}  +{i}{\bf A}\right)\psi_i^*\cdot (M_{red}^{*})^{-1}\cdot\left(\frac \nabla {\kappa}  -{i}{\bf A}\right)\psi_i- |\psi_i|^2 +\frac{1}{2} |\psi_i|^4\right] \right. \nonumber \\ 
\left. +2\tilde b\bm{J}_{\uparrow}\cdot (M_{\rm red}^{*})^{-1}\cdot\bm{J}_{\downarrow}+\tilde \mu(|\psi_{\uparrow}|^2-|\psi_{\downarrow}|^2)B_{||}+|\nabla\times {\bf A}-{\bf B}_{\rm ext}|^2\right\}.
\eea
Here $\bm{J}_i={\rm Re}\{-i\psi_i^* (\kappa^{-1}\nabla-i\bm{A})\psi^*\}$ is the dimensionless current, and $\tilde b$ and $\tilde \mu$ are dimensionless parameters corresponding to the current-current coupling $b$ in (\ref{eq:CCcoupling}) and $g\mu_B$ in (\ref{eq:KSP}),
respectively. The parameter $\kappa\approx 2.3 $ is the ratio of the in-plane penetration depth to the in-plane coherence length. 
Also 
\be
M_{\rm red}^{*}=\text{diag}(1,1,m^*_{\perp}/m^*_\parallel).
\ee
To find the magnetic field  and condensate configurations that minimize  the free energy functional  (\ref{EQ:finalFreeEnergy})  we used  the commercial finite-element method solver COMSOL.

\begin{figure}[htb]
\centering
\includegraphics[width=\linewidth]{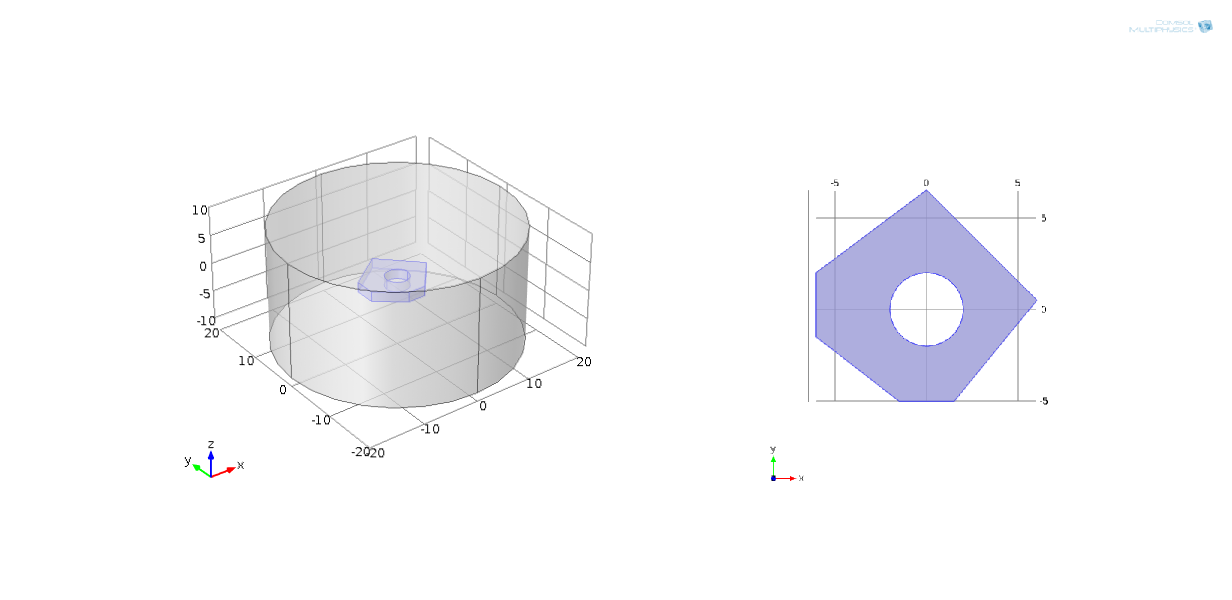}
\caption{On the left is shown a representation of the simulated geometry. On the right is a top-down view of the simulated ring. The axes in both figures are labelled in units of $\lambda_{||}$. }
\label{fig:geometry}
\end{figure}

We will focus on the results for the  single annular sample  shown in   figure \ref{fig:geometry}. The ring's inner radius is 2 $\lambda_{||}$ while the outer radius averages 5.29 $\lambda_{||}$ where $\lambda_{||}(T=0)$ is given as $152$ nm for Sr$_2$RuO$_4$ in \cite{maeno_review}. The height of the sample is 1.7 $\lambda_{||}$. This geometry is designed to approximate the  experimental sample shown in Figure [\ref{fig:ringAndMoment}]. The  ring is centered  in a cylindrical volume $\Omega$ of height 20 $\lambda_{||}$ and radius 20 $\lambda_{||}$. The dimensions of the cylinder were chosen to balance the competing effects of increased computation time for larger cylinders against  spuriously high magnetic field  energies  caused by confining the field in too small a volume.

The externally imposed magnetic field  ${\bf B}= (B_x,B_y,B_z)$  is established  by imposing the inhomogeneous Dirichlet boundary condition 
\bea \label{EQ:ABoundaryConditions}
A_x &=& \frac 12 (B_y z-B_z y),\nonumber\\
 A_y &=& \frac 12 (B_z x-B_x z), \nonumber\\
 A_z&=& \frac12 ( B_x y- B_y x).
\eea
on the surface  the $\partial \Omega$ of the cylinder.

The boundary conditions we impose on  the order-parameter fields  in 
(\ref{EQ:finalFreeEnergy}) are the ``natural''  boundary conditions that arise from the variational problem of free energy minimization.  In other words we  require the vanishing of the integrated out  variation terms on the surface of the superconductor. This leads to    
\bea
{\bf n}\cdot (M_{\rm red}^*)^{-1}\cdot\left(\frac{\nabla}{\kappa} -i{\bf A}-\frac{\tilde \beta}{\kappa}\bm{J}_{\downarrow}\right)\psi_{\uparrow} &=&0\nonumber\\ 
{\bf n}\cdot (M_{\rm red}^*)^{-1}\cdot\left(\frac{\nabla}{\kappa} -i{\bf A}-\frac{\tilde \beta}{\kappa}\bm{J}_{\uparrow}\right)\psi_{\downarrow} &=&0
\eea
on the surface of the superconducting ring.

We find the  fields that minimize the free energy   by first choosing initial conditions for the magnetic vector potential and for the order-parameter fields, and then allowing them to relax to equilibrium.   As the applied magnetic field along the z-axis is increased, we see vortices enter the superconductor and increase winding numbers about the ring. Each entry  results   in a discontinuous step in the magnetization. 
These transitions turn out to be  difficult to model reliably. Although  vortices enter the sample,  the regions of metastabilty are large and geometry dependent. 
We therefore adopted a strategy that  is   similar to the field-cooling used in the actual experiments. We begin by imposing   initial conditions 
\be
\psi_{\uparrow}=\exp[in_{\uparrow}\phi];\quad\psi_{\downarrow}=\exp[in_{\downarrow}\phi]
\ee
that correspond  to a selected  flux state. (Here, $\phi$ is the azimuthal angle around the ring. For instance, $(n_{\uparrow},n_{\downarrow})=(0,0)$ corresponds to the zero flux state while $(1,0)$ and $(0,1)$ correspond to half flux 
states.) 
After selecting  the desired  winding numbers,  and setting the externally imposed magnetic field,  we  allow  the system to relax to a local minimum of the free energy. This free energy is then calculated by evaluating  the integral in (\ref{EQ:finalFreeEnergy}). In this way we are  able to construct diagrams of  free energy versus applied magnetic field in the z-direction for different flux states and for different values of in-plane magnetic field. These diagrams reveal which flux state is energetically  favourable at each value of applied z-axis field. Using these plots, we numerically compute  the derivatives of the free energy with respect to the applied z-axis field, and thus construct the magnetization curves.

\section{Results}
\label{SEC:results}

\begin{figure}[htb]
\centering
\includegraphics[width=1.0 \textwidth]{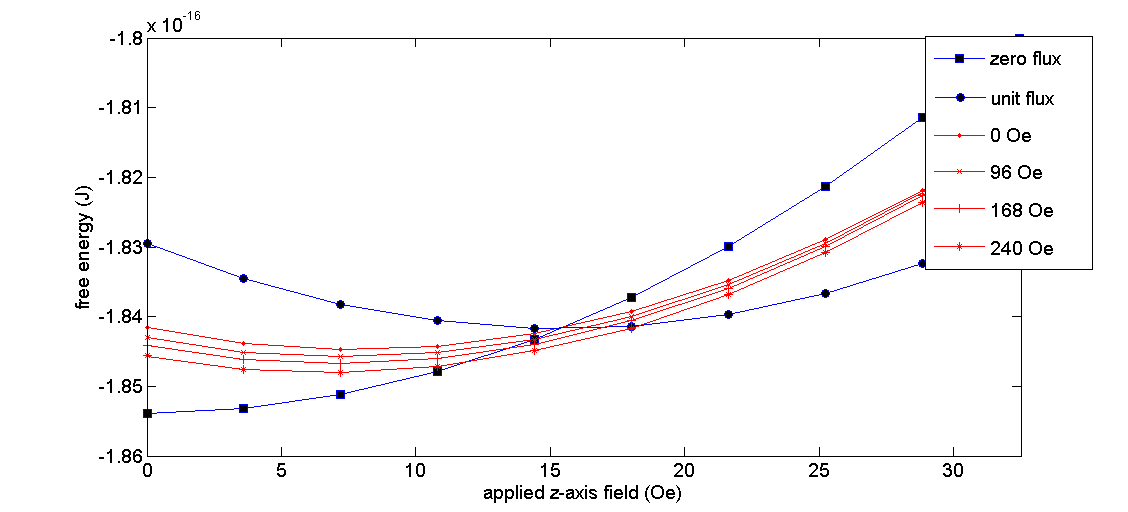}
\caption{This figure shows an example of calculated relative free energies of the integer and half-flux states near the first transition. Each red line represents the free energy of the half-flux state at different values of applied in-plane magnetic field. Higher magnetic fields are seen to result in relatively lower free energies of the half-flux state.}
\label{fig:freeEnergy}
\end{figure}

Figure \ref{fig:freeEnergy} illustrates  the free energy of the system versus applied z-axis field (from 0 to 32.5 Gauss), and  for a few values of applied in-plane magnetic field. The blue line depicts the (0,0) and (1,1) integer-flux states while the red line depicts the (1,0) half-flux state.\footnote{Applying an in-plane field produces a spin polarization for both both integer and half integer flux states. We have  plotted the data so as to show the only the {\it relative} free energy of the integer and half integer states. We did this by  shifting  all curves by an in-plane-field dependent  constant so that the integer-flux state's data overlap one another.} One can see from the diagram that the half-flux state has a reduced free energy versus the integer-flux states for higher values of in-plane field.

\begin{figure}[htb]
\centering
\includegraphics[width=1.0 \textwidth]{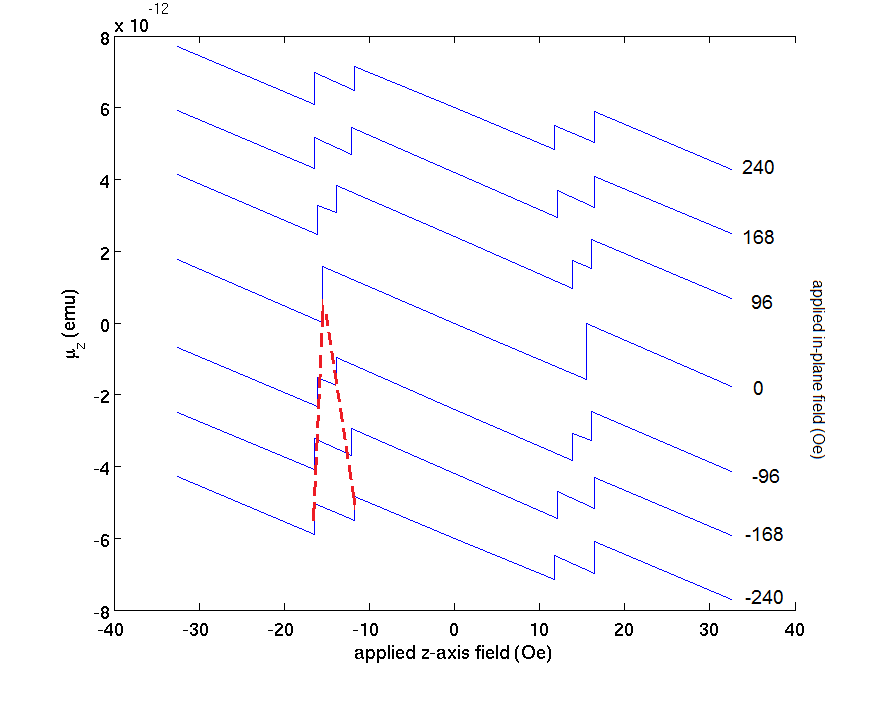}
\caption{Computed magnetization curves obtained  using $\rho_{sp}/\rho_s=0.25$ and a magnetic dipole moment of 0.55$\mu_B$.}
\label{fig:magCurves}
\end{figure}

Applying a parabolic fit to the free energy curves, one can take the the derivative with respect to the applied magnetic field to obtain magnetization curves. Figure \ref{fig:magCurves} displays magnetization curves versus z-axis field for various values of in-plane field as in Figure 2. The magnetization curves are seen to be qualitatively similar. By adjusting $\tilde b$ and $\tilde \mu$ one may obtain the desired minimum in-plane stabilization field and stability region growth rate. 

The dimensionless current-current coupling parameter $\tilde b$ can be related to the ratio of the spin and charge fluid densities via 
\begin{equation}
	\frac{\rho_{sp}}{\rho_s}=\frac{1-\tilde b}{1+\tilde b}.
\end{equation}
Typical values of $\tilde b$ which matched the experimental data were 0.4-0.6 corresponding to $\rho_{sp}/\rho_s$ between 0.25 and 0.43.

If $\tilde \mu$ is interpreted as the magnetic dipole moment per particle of the spin condensate, it gives a magnetic dipole moment in units of $\sqrt{2}e^2 B_{c}\lambda^2m_e^{-1}$. Using $B_{c}=230$ Oe and  $\lambda=152$ nm, the value of this unit is approximately $2.12\times10^{-23}J/T\approx 2.2\mu_B$. Typical values of the magnetic dipole moment which fit the data were found to be around 0.6 $\mu_B$.

\begin{figure}[htb]
\centering
\includegraphics[width=1.0 \textwidth]{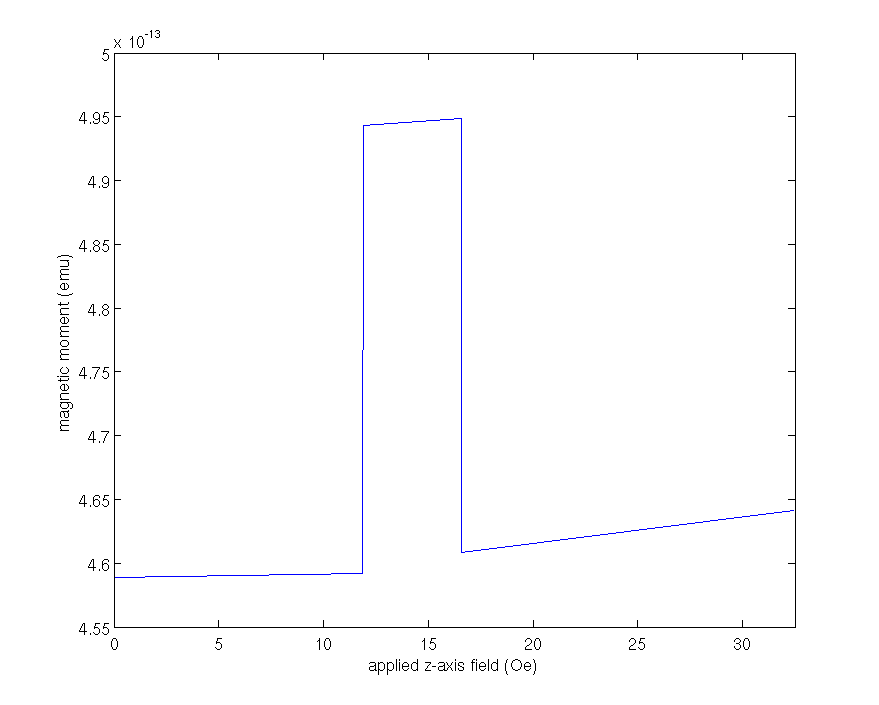}
\caption{The calculated in-plane spin magnetic moment for 200 Oe of applied in-plane field.}
\label{fig:inPlaneMu}
\end{figure}

The magnitude of the in-plane spin magnetic moment for the case of 240 Oe in-plane field is plotted in Figure \ref{fig:inPlaneMu}. Near zero applied z-axis field the system is in the $(n_{\uparrow},n_{\downarrow})=(0,0)$ state. The presence of the in-plane magnetic field induces an in-plane magnetic moment even in the absence of kinematic spin polarization.Thus, there is a non-zero moment even in the integer flux states. The presence of kinematic spin polarization accounts for the sudden increase in moment at the transition to the half-flux state near 12 Oe. This additional moment vanishes when the system transitions to the unit fluxoid state at 16 Oe.

\section{Discussion}
\label{SEC:discussion}

We have shown that the model described in section II can qualitatively reproduce the results of the experiment described in \cite{budakian}. Using kinematic spin polarization and a current coupling term in a two-component Ginzburg-Landau model, numerical results show that an in-plane field results in the increasing stability of a half-quantum vortex state. This leads to a wedge-like stability region in the half-flux state versus applied c-axis field as shown in Figure \ref{fig:magCurves}. Moreover, this is achieved using physically reasonable values of $\mu$ and $\rho_{sp}/\rho_s$. 

In comparing Figure \ref{fig:magCurves} to Figure \ref{fig:momentData} there are some disrepencies that should be acknowledged. Firstly, the periodicity of the numerical result is approximately 30 Oe while that of the data is 16 Oe. This is due to the fact that, assuming a penetration depth of 152 nm, the simulated ring's hole diameter is 0.6 $\mu$m while the actual sample has a hole diameter of 0.75 $\mu$m. A smaller hole necessitates a larger applied field in order to achieve the flux necessary to induce a fluxoid transition.

A more perplexing difference is in the magnitude of the moments, the numerical result's moments being an order of magnitude larger than those shown in Figure \ref{fig:momentData}. These small magentic moments seem to be found only in the smaller samples examined by Jang, et. al. The moment data collected from larger samples, as in Figure 2 of \cite{budakian2}, seems to be of the appropriate order of magnitude. At the present time, we can only speculate that the small moments are due to weakening of the superconductivity caused by crystal defects incurred during fabrication of the rings.

This discrepency in the magnitudes of the magnetic moments makes the interpretation of the experiments in \cite{budakian} difficult. One possible scenario explaining the half-height jumps in magnetization were Abrikosov vortices piercing the side-wall nearly half-way between the top and bottom of the ring. One of the several arguments against this was that the magnitude of the induced spin magnetic moment $\mu_{HI}$ is an order of magnitude less than that produced by a side-wall vortex. The estimates of $\mu_{HI}$ were determined by Jang, et. al. by using the formula $\mu_{HI}=\delta H_z\Delta\mu_z/4(H_x-H_{x,min})$ where $\Delta\mu_z$ is the jump in magnetic moment upon entry of a unit vortex, $\delta H_z$ is the width of the stability wedge, and $H_x-H_{x,min}$ is the amount of applied in-plane field over the minimum necessary to see half-flux states. For the data shown in Figure \ref{fig:momentData}, the implied $\mu_{HI}\approx 9\times 10^{-15}$ emu while, from Figure \ref{fig:inPlaneMu}, $\mu_{HI}\approx 3.5\times 10^{-14}$ emu. So, since the measured $\Delta\mu_z$  and $\mu_{HI}$ appear to be poorly understood this argument loses some credibility. A better understanding will require an understanding of the character and stability of wall-vortex states.

\section{Acknowledgments} 
We would like to acknowledge useful discussions with Anthony Leggett, David Furgeson, and Victor Vakaryuk. This work was supported by the National Science Foundation under grant DMR 09- 
03291.

 \end{document}

%% file: mma_defs.tex
\def\bi{\begin{itemize}}
\def\ei{\end{itemize}}
\def\bea{\begin{eqnarray}} 
\def\eea{\end{eqnarray}}
\def\be{\begin{equation}}
\def\ee{\end{equation}}
\def\line{\hbox to \hsize}    
\def\frac #1#2{{#1\over #2}}

\def\1{\mbox{\bf I}}

\def\bm#1{\mbox{\boldmath$#1$}} 

\newenvironment{Quote}
{\begin{list}{}{%
\setlength{\leftmargin}{10 pt}
\setlength{\rightmargin}{\leftmargin}}
\item[]}
{\end{list}}

     {\refstepcounter{exercisenumber} \begin{Quote}\small{\sl
     Exercise~\thechapter.\theexercisenumber\/}:}
{\end{Quote}}

{\begin{Quote}\small{\sl Example\/}:}
{\end{Quote}}

\def\levelonelist{
        \begin{list}{\mybulA}%
                        {
        \setlength{\topsep}{0pt}
        \setlength{\parsep}{0pt}
        \setlength{\partopsep}{0pt}
        \setlength{\itemsep}{0pt}
                        }
                }
\def\leveltwolist{
        \begin{list}{\mybulB}%
                        {
        \setlength{\topsep}{0pt}
        \setlength{\parsep}{0pt}
        \setlength{\partopsep}{0pt}
        \setlength{\itemsep}{0pt}
                        }
                }

\def\el{\end{list}}